\documentclass[a4paper,11pt]{article}
\pdfoutput=1 
\usepackage{jheppub} 
\usepackage[T1]{fontenc} 

\usepackage{CJK}
\usepackage{graphicx}
\usepackage{multirow}

\usepackage{soul}

\usepackage{lineno}

\title{Search for coherent elastic neutrino-nucleus scattering at a nuclear reactor with CONNIE 2019 data}

\author[a]{Alexis Aguilar-Arevalo,}
\author[b]{Javier Bernal,}
\author[c]{Xavier Bertou,}
\author[d,n]{Carla Bonifazi,}
\author[e]{Gustavo Cancelo,}
\author[d]{Victor G. P. B. de Carvalho,}
\author[a]{Brenda A. Cervantes-Vergara,}
\author[b]{Claudio Chavez,}
\author[f]{Gustavo Coelho Corr\^ea,}
\author[a]{Juan C. D'Olivo,}
\author[g]{Jo\~ao C. dos Anjos,}
\author[e]{Juan Estrada,}
\author[h]{Aldo R. Fernandes Neto,}
\author[e,i]{Guillermo Fernandez Moroni,}
\author[d]{Ana Foguel,}
\author[e]{Richard Ford,}
\author[j]{Juli\'an Gasanego Barbuscio,}
\author[b]{Juan Gonzalez Cuevas,}
\author[e]{Susana Hernandez,}
\author[k]{Federico Izraelevitch,}
\author[l]{Ben Kilminster,}
\author[e]{Kevin Kuk,}
\author[g]{Herman P. Lima Jr,}
\author[g,n]{Martin Makler,}
\author[a]{Mauricio Martinez Montero,}
\author[d]{Larissa Helena Mendes,}
\author[b]{Jorge Molina,}
\author[g]{Philipe Mota,}
\author[d,1]{Irina Nasteva,\note{Corresponding author.}}
\author[i]{Eduardo Paolini,}
\author[m]{Dario Rodrigues,}
\author[a]{Y. Sarkis,}
\author[c,e]{Miguel Sofo Haro,}
\author[b]{Diego Stalder,}
\author[e]{Javier Tiffenberg}

\affiliation[a]{Universidad Nacional Aut\'onoma de M\'exico, Distrito Federal, M\'exico}
\affiliation[b]{Facultad de Ingenier\'ia, Universidad Nacional de Asunci\'on, Paraguay}
\affiliation[c]{Centro At\'omico Bariloche and Instituto Balseiro, Comisi\'on Nacional de Energ\'ia At\'omica (CNEA),  Consejo Nacional de Investigaciones Cient\'ificas y T\'ecnicas (CONICET), Universidad Nacional de Cuyo (UNCUYO), San Carlos de Bariloche,  Argentina}
\affiliation[d]{Instituto de F\'isica, Universidade Federal do Rio de Janeiro, Rio de Janeiro, RJ, Brazil}
\affiliation[e]{Fermi National Accelerator Laboratory, Batavia, IL, United States}
\affiliation[f]{Eletronuclear, Angra dos Reis, RJ, Brazil}
\affiliation[g]{Centro Brasileiro de Pesquisas F\'{\i}sicas, Rio de Janeiro, RJ, Brasil}
\affiliation[h]{Centro Federal de Educa\c{c}\~ao Tecnol\'ogica Celso Suckow da Fonseca, Angra dos Reis, RJ, Brazil}
\affiliation[i]{Instituto de Inv. en Ing. El\'ectrica ``Alfredo Desages'' (IIIE), Dpto. de Ing. El\'ectrica y de Computadoras, CONICET and Universidad Nacional del Sur (UNS), Bah\'ia Blanca, Argentina}
\affiliation[j]{Departamento de  F\'isica, FCEN, Universidad de Buenos Aires, Buenos Aires, Argentina}
\affiliation[k]{Instituto Dan Beninson, Universidad Nacional de San Mart\'in (UNSAM), CNEA and CONICET, Argentina}
\affiliation[l]{Universit\"at Z\"urich Physik Institut, Zurich, Switzerland}
\affiliation[m]{Departamento de  F\'isica, FCEN, Universidad de Buenos Aires and IFIBA, CONICET, Buenos Aires, Argentina}
\affiliation[n]{International Center for Advanced Studies \& Instituto de Ciencias F\'isicas,  ECyT-UNSAM and CONICET, Buenos Aires, Argentina}

\emailAdd{Irina.Nasteva@cern.ch}

\abstract{The Coherent Neutrino-Nucleus Interaction Experiment (CONNIE) is taking data at the Angra 2 nuclear reactor with the aim of detecting the coherent elastic scattering of reactor antineutrinos with silicon nuclei using charge-coupled devices (CCDs). 
In 2019 the experiment operated with a hardware binning applied to the readout stage, leading to lower levels of readout noise and improving the detection threshold down to 50 eV\@.
The results of the analysis of 2019 data are reported here, corresponding to the detector array of 8 CCDs with a fiducial mass of 36.2~g 
and a total exposure of 2.2~kg-days.
The difference between the reactor-on and reactor-off spectra shows no excess at low energies and yields upper limits at 95\% confidence level for the neutrino interaction rates. In the lowest-energy range, $50-180$~eV, the expected limit stands at 34 (39) times the standard model prediction, while the observed limit is 66 (75) times the standard model prediction with Sarkis (Chavarria) quenching factors.}

\begin{document}
\maketitle


\section{Introduction}

Coherent elastic neutrino-nucleus scattering (CE$\nu$NS) is a standard model process in which a neutrino scatters elastically off a nucleus in a coherent way~\cite{Freedman:1973yd}. The enhanced cross-section due to the neutrino interaction with the entire nucleus makes this process detectable with smaller detectors, as long as they have a low energy threshold for nuclear recoils. 
CE$\nu$NS provides a new window into the low-energy neutrino sector and the interest in its detection has been growing as a potential probe for new physics~\cite{Papoulias:2019}.

The CE$\nu$NS differential cross-section for the coherent elastic scattering of antineutrinos off a nucleus at rest is 
\begin{equation}
\label{eq:crossSM}
\frac{d\sigma_{SM}}{dE_R}\left(E_{\bar{\nu}_e}\right)=\frac{G_F^2}{8\pi}Q_W^2\left[2-\frac{2E_R}{E_{\bar{\nu}_e}}+\left(\frac{E_R}{E_{\bar{\nu}_e}}\right)^2-\frac{ME_R}{E_{\bar{\nu}_e}^2}\right]M\vert F(q)\vert^2\,,
\end{equation}
with the weak charge
\begin{equation}
\label{eq:QW}
Q_W = N - (1 -4 \sin^2\theta_W)Z\,,
\end{equation}
where $Z$($N$) is the number of protons (neutrons), $M$ the mass of the nucleus, $G_F$ the Fermi coupling constant, $E_{\bar{\nu}_e}$ the antineutrino energy, $E_R$ the nuclear recoil energy, and $F(q)$ is the nuclear form factor.

There are currently two  approaches for the observation of CE$\nu$NS. The first, which uses a stopped pion beam generating neutrinos with energies of $\sim$20\,MeV, presents compelling advantages for the detection of coherent scattering. The neutrinos are produced at the highest energies where the coherence of the process is maintained, giving the largest possible nuclear recoil energies for CE$\nu$NS\@. In addition, the timing structure of the beam allows for strong background suppression. The COHERENT Collaboration has successfully observed CE$\nu$NS for the first time from a stopped pion beam at SNS using a CsI[Na] detector~\cite{Coherent}, and more recently with a LAr detector~\cite{COHERENT:2020iec}. 

The second approach is based on the MeV-energy neutrinos produced in nuclear reactors~\cite{TEXONO:2005fmk}. These lower-energy neutrinos generate smaller recoils in the detector material compared to stopped pion beams, and in this case there is no timing information that could be used to suppress background in the case of fast detectors. 
Also, because the reactor neutrino spectrum peaks at low energies ($\sim$1 MeV), the neutrino flux that can be accessed depends on lowering the detection threshold.
The detection of coherent scattering from reactor neutrinos is experimentally more challenging but has some unique features as a probe for new physics in the low-energy neutrino sector. The lower momentum exchange in CE$\nu$NS from reactor neutrinos eliminates the dependence on the nuclear form factor, with $F(q) \sim 1$ in Eq.~\ref{eq:crossSM}, and enhances the sensitivity to potential effects predicted in new physics scenarios, such as an anomalous magnetic moment~\cite{MagneticMoment} or millicharge~\cite{Millicharge} of the neutrino, light sterile neutrinos~\cite{Kosmas:2017zbh}, weak mixing angle~\cite{WeakAngle, WeakMixingAngle}, and non-standard interactions of neutrinos such as light mediators~\cite{Proceedings:2019qno,Farzan:scalar}.

Several experiments are either running or in a preparation stage with the aim of observing for the first time CE$\nu$NS with reactor neutrinos \cite{CONUS:2020skt, MINER, TEXONO:2005fmk, RED:2012hpm}. 
The Coherent Neutrino-Nucleus Interaction Experiment (CONNIE) has so far been the most sensitive at low energies.
Operating with a detector of charge-coupled devices (CCDs) at the Angra 2 nuclear reactor in Brazil, CONNIE has demonstrated with the results of its analysis of 2016--2018 data~\cite{connie:2019} the sensitivity to new physics by establishing competitive limits for light mediators at the lowest mediator masses~\cite{connie:LM}. 
In this work we present the results of the CONNIE experiment based on data collected in 2019, obtained with a lower energy threshold thanks to a different readout strategy, in which a hardware binning in one direction increases the signal-to-noise ratio. 
In addition, several new calibration tools and analysis methods were developed to quantify and improve the detector performance.

This paper is organised as follows. Section~\ref{sec:CONNIEexp} describes the CONNIE experiment, CCD readout strategy and method. Section~\ref{sec:calib} presents the calibration and performance of the experiment.
Backgrounds are discussed in Section~\ref{sec:backg}. The selection of the neutrino candidate events is presented in Section~\ref{sec:neutinoevents}. 
The search for CE$\nu$NS and the measurement results are presented in Section~\ref{sec:onoff}, while the conclusions and outlook are discussed in Section~\ref{sec:conclusion}.

\section{The CONNIE experiment and data set}\label{sec:CONNIEexp}

The CONNIE detector consists of 12 fully depleted high-resistivity silicon scientific CCDs, mounted horizontally in a tower inside a copper cold box, surrounded by passive shielding. 
Each CCD sensor is an array of $4120\times 4120$ square pixels of $15\times 15~\mu$m$^2$ area each and 675~$\mu$m thickness, with a mass of 6.0~g. 
The CCDs operate at cryogenic temperatures and are contained inside a vacuum vessel, shielded with 15~cm of lead, sandwiched between two layers of 30~cm of polyethylene. 
The CONNIE experiment is operating since 2016 at a distance of about 30\,m from the core of the 3.95\,GW thermal power Angra 2 nuclear reactor,  in a shipping container placed just outside its containment dome. A more detailed technical description of the experiment, image processing and event extraction is presented in Refs.~\cite{connie:2019, connie:2016}. Here we summarise the main features and focus on the new readout method for the sensors.

A search for CE$\nu$NS events is performed by comparing the energy spectra of data between the periods with the reactor operating at full power (reactor on) and during the scheduled shutdown of about 1 in every 13 months (reactor off). 
The current data set covers 31.85 days of operation with the reactor on and 28.25 days with the reactor off, collected in 2019 with 8 CCDs that show stable operation and good quality data, while the remaining 4 CCDs were excluded from the analysis due to high noise or charge transfer inefficiency. 
The experiment active fiducial mass after acceptance effects from geometrical and event size cuts (as detailed in Section~\ref{sec:onoff}) is 36.24~g, giving a total exposure of 2.2 kg-days.

\subsection{CCD operation and readout}\label{sec:ccd}

The CONNIE CCD sensors were developed by the experiment in collaboration with the LBNL Micro Systems Labs~\cite{LBNLMSL}, as a spin-off from the fully depleted thick detectors designed for astronomical instruments such as DECam \cite{DES:2015wtr} and DESI~\cite{DESI:2018kpn}.
The CCD thickness was increased to 675~$\mu$m, making them the thickest CCDs ever fabricated. 
The CCDs are mounted horizontally in a copper box, which is kept inside a copper vacuum vessel ($10^{-7}$ torr). The sensors orientation is such that the pixel gates are on their front (or top) side, and the substrate depletion voltage of 70~V is applied to the back (bottom) side.
In order to reduce the thermally-generated dark current in the silicon, the sensors are cooled to temperatures below 100~K by means of a closed-cycle helium cryocooler.

Each CCDs is read out sequentially, by moving the charge of each pixel  towards one corner of the sensor where it is read out via a single output amplifier.  
More pixel values are read beyond the physical extent of the CCD, forming the so-called overscan region, used to monitor the baseline of the readout electronics.
At the same clock rate as the readout amplifier,
a second output of each sensor is read out via a second amplifier,  without any charge from the array, in order to generate a pure noise image used to monitor the correlated noise of the system. 
The image formed by the array of charges is processed by subtracting the mean value of the overscan image, then subtracting the master bias image formed from the median of a large number (60) of images, and subtracting the correlated noise obtained from the second amplifier readout. 
More details on the image processing can be found in Ref.~\cite{connie:2019}.

\subsection{CCD readout mode with hardware binning}\label{sec:DAQ}

A new hardware binning mode of data taking was implemented in 2019, which is described as follows.  
By controlling the voltages applied to the phases of the three pixel gates called vertical clocks, each pixel develops a potential well that stores charges. 
The readout of the CCD is done sequentially, one pixel at a time, by phasing the vertical clocks in an alternate sequence, the charge in the two dimensional array is first moved into a special register called the horizontal register. 
Since the vertical clocks move the charge towards the horizontal register, that register receives the charge from the last row in the array.  
The entire row of pixels is read out by shifting the charge in the horizontal register serially through the amplifier by phasing another set of three voltages, the horizontal clocks.  The horizontal register can also be used as an accumulator, a mode of acquisition that we call hardware binning.  The vertical clocks in the array can be applied cyclically to move charges from subsequent pixels into the horizontal register. 
Each pixel in the horizontal register then accumulates the summed charge for as long as its capacity does not overflow.  At the end of the hardware binning sequence the horizontal register is read out.

\begin{figure}[tb]
\centering
\includegraphics[width=0.4\textwidth]{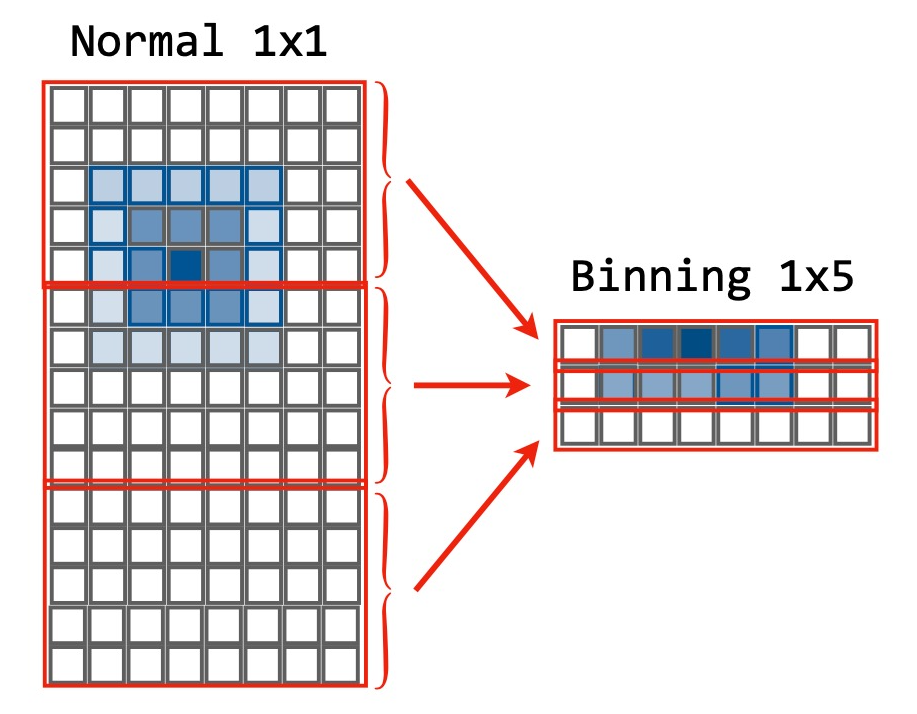}
\includegraphics[width=0.59\textwidth]{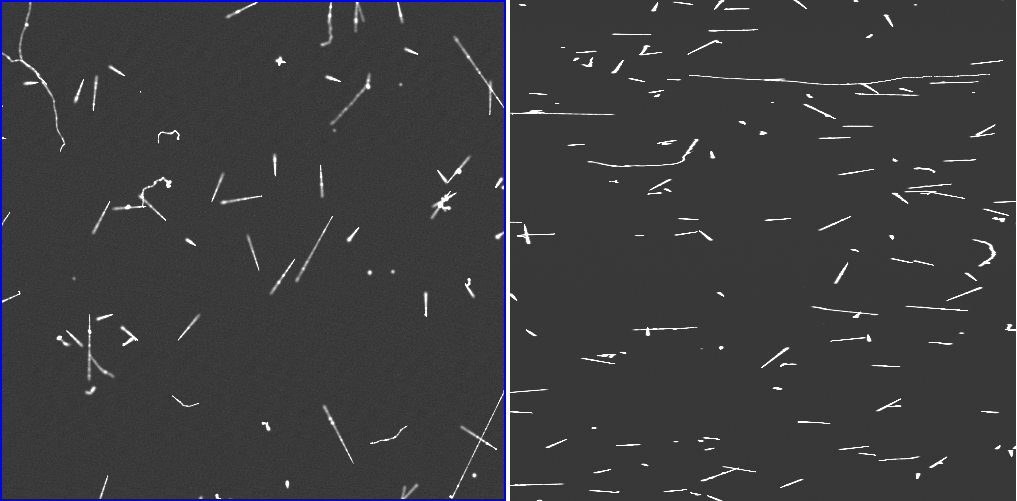}
\caption{A diagram of hardware binning, in which the charges of 5 vertical pixels are added up and read together (left), and a part of a standard image (middle) compared to one taken with the hardware binning (right). The binned image appears compressed in the vertical direction and the full image has 5 times fewer pixels in this direction.}
\label{fig:binning}
\end{figure}

The binning technique is used to minimise the readout noise, which is added by the readout amplifier. Since the charge of $N$ pixels is added up before reading, the effective readout noise per pixel is $N$ times smaller. 
The price paid for using binning is loss of pixel resolution in the direction of the binning, as each pixel represents a silicon area $N$ times larger in size. 
Different binning schemes and image exposure times were studied, in order to have low readout noise while also keeping a low pixel occupancy due to background. 
The optimal readout strategy was found to be with a binning of $N=5$ in the vertical direction and exposure times of 1 hour per image. Pixels therefore have an effective size of 15~$\mu$m by 75~$\mu$m (see Fig.~\ref{fig:binning}).

\subsection{Strategy to search for CE$\nu$NS }\label{sec:ana}

Several new tools and calibration techniques have been employed in the analysis in this work. 
They were developed using the reactor-off data that provide only the radioactive and cosmogenic background and sensor performance information, without any neutrino events expected. 
A blind analysis was performed, 
without looking at reactor-on data, in order not to bias the selection with respect to a possible signal.

The calibration techniques and selection tools were developed using a combination of reactor-off data and simulations. Energy calibration is performed using fluorescence peaks in the data (Section~\ref{sec:gain_calibration}), and depth calibration is derived from the width of muon tracks (Section~\ref{sec:size_vs_depth_calibration}).
The background data sample was then studied in the low-energy range (below around 1~keV), where neutrino interactions are expected, and at higher energies.
On-chip noise sources were quantified in the reactor-off data and a study of fake events that can mimic a neutrino signal at very low energies was conducted. The evaluation of the uncertainties and contributions of spurious events is critical for the sensitivity to low-energy events in the region of interest for reactor neutrinos, and special care is therefore needed. 
The event size information was found to be a very efficient way to remove defective tracks that do not fit into a single hit in the CCD (Section~\ref{sec:lowenergy}).

After this, the neutrino selection criteria (Section~\ref{sec:neutinoevents}) are defined using samples of simulated neutrino events, as in the previous analysis~\cite{connie:2019}. 
Neutrino scattering events are generated with a uniform probability in the active volume 
and a uniform distribution in energy, and are then added to raw images from the reactor-off data set and reconstructed in the same way as data. 
Consistency checks are applied to the events passing the selection criteria.

Finally, the reactor-on data sample is unblinded.
The rates of the high-energy background events are compared between the reactor-on and off periods to check detector stability.
The selection criteria determined previously are applied to the low-energy events in the reactor-on period and sanity checks are again performed on these data.
Then, in order to compute the neutrino rate, the total event rate during the shutdown period of the detector is subtracted from the total event rate measured with the reactor running (Section~\ref{sec:onoff}).

\section{Detector calibration and performance}\label{sec:calib}

The techniques and tools to calibrate the sensor and to measure key performance parameters were updated with respect to the previous analysis of 2016--2018 data~\cite{connie:2019}. 
The following subsections detail these new tools and their use in the 2019 data analysis.

\subsection{Gain calibration}
\label{sec:gain_calibration}

\begin{figure}[tb]
\centering
\includegraphics[width=0.5\textwidth]{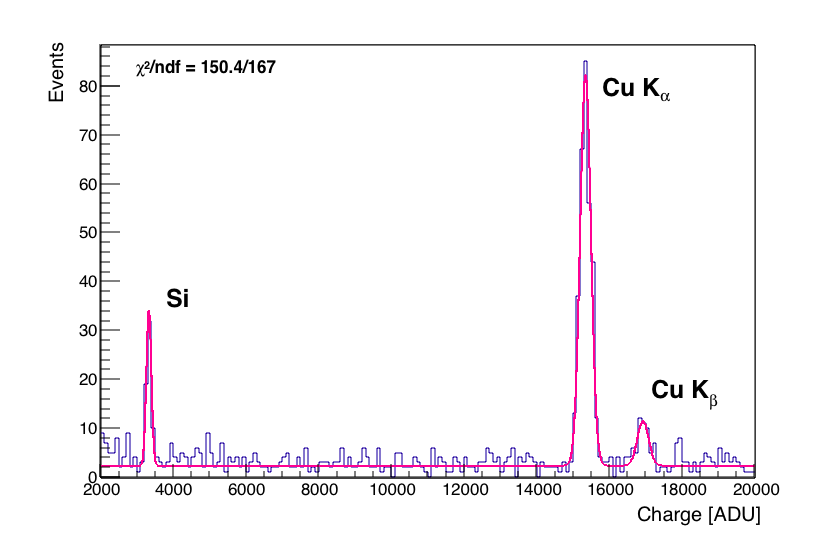}
\includegraphics[width=0.47\textwidth]{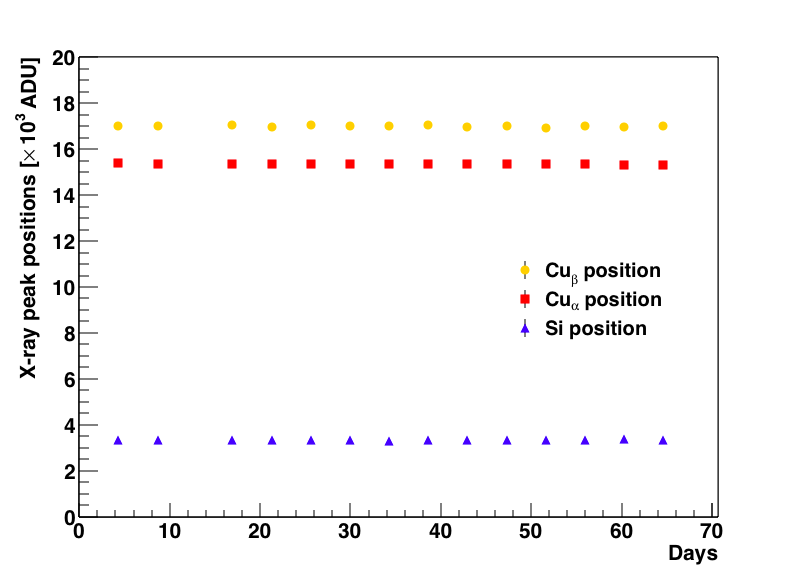}
\caption{(Left) Energy spectrum from 100 images taken from the edge regions of one CCD\@, superimposed with the fit to the main fluorescence x-ray peaks. (Right) Time evolution of the three x-ray peak positions of the same CCD during the data taking period.}
\label{fig:calibration}
\end{figure}

The energy, or gain, calibration of the CCDs is based on the emission of copper and silicon fluorescence x-rays from excitation of cosmogenic origin or due to the natural radioactivity of the surrounding materials. 
Because of the silicon CCD substrates and copper structures surrounding them, the  $\rm{Cu}_{\alpha}$ (8.047 keV),  $\rm{Cu}_{\beta}$ (8.905 keV) and Si (1.740 keV) emissions are readily observed as peaks in the energy spectrum.
The calibration procedure was expanded with respect to the previous analysis to include information from all three peaks, and was applied to samples of 100 consecutive images, corresponding to around four days of data taking, for each CCD\@. 
The peak positions are determined from a fit to the energy spectrum of events from the edges of the CCD, where most fluorescence events lie, described by a function that is the sum of three Gaussians for the peaks and a constant background term. 
An example energy spectrum with the fit overlaid is shown in Fig.~\ref{fig:calibration}. 
The Gaussian means, width of the $\rm{Cu}_{\alpha}$ peak, and all normalisations are free to vary in the fit, while the widths of the other two Gaussians are determined from the one of the $\rm{Cu}_{\alpha}$ peak, assuming a constant Fano factor and Poisson charge carrier distributions. 

To obtain the gain, the peak positions are fitted with a second-order polynomial with a zero constant term, in which the linear term is the reciprocal of the gain and the small quadratic term reflects possible non-linearity. 
The maximum non-linearity for all CCDs was found to be smaller than 1\% and non-linearity effects were therefore discarded as negligible within the low-energy range considered. 
The validity of using 100 images was cross-checked by applying the calibration to smaller groups of images and showed variations compatible with the limited sample sizes. 
The gain stability is also shown in Fig.~\ref{fig:calibration} for one CCD and the whole data taking period. 
All CCD gains are found to be stable with time.

\subsection{Size versus depth calibration}
\label{sec:size_vs_depth_calibration}

\begin{figure}[tb]
    \centering
    \includegraphics[width=0.7\textwidth]{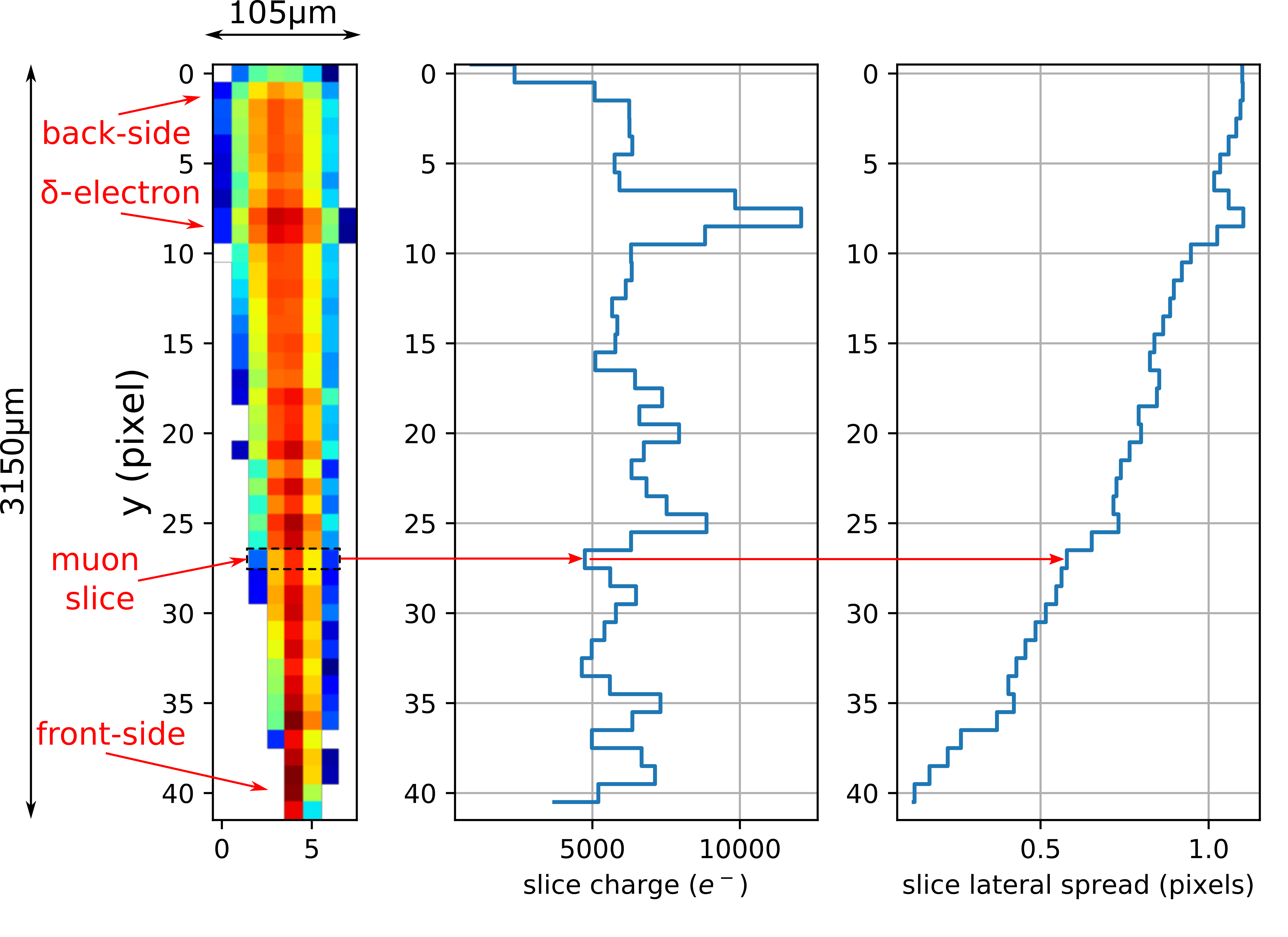}
    \caption{Image of a y-axis muon. For details see text.}
    \label{fig:muon1x5}
\end{figure}

The depth calibration is essential to determine the detection efficiency and to identify fake events that may mimic neutrino interactions. 
The calibration curve relates the lateral spread (or size) of low-energy events with their interaction depth in the CCD silicon substrate. 
Although the holes are initially produced inside the volume of a single pixel, while they drift to the collection wells of the pixels they diffuse and by the time they reach the potential well on the front, they spread over a few pixels~\cite{haro2020studies}. 
Holes produced close to the CCD back side have more time to diffuse before being collected, and therefore, they spread more than the holes produced close to the pixel collection wells. 
Given the fact that each pixel is affected by readout noise, those events with the holes distributed over a few pixels become less likely to be detected than events with only one pixel once a threshold is applied to extract them, due to their lower signal-to-noise ratio. 

\begin{figure}[tb]
    \centering
    \includegraphics[width=0.65\textwidth]{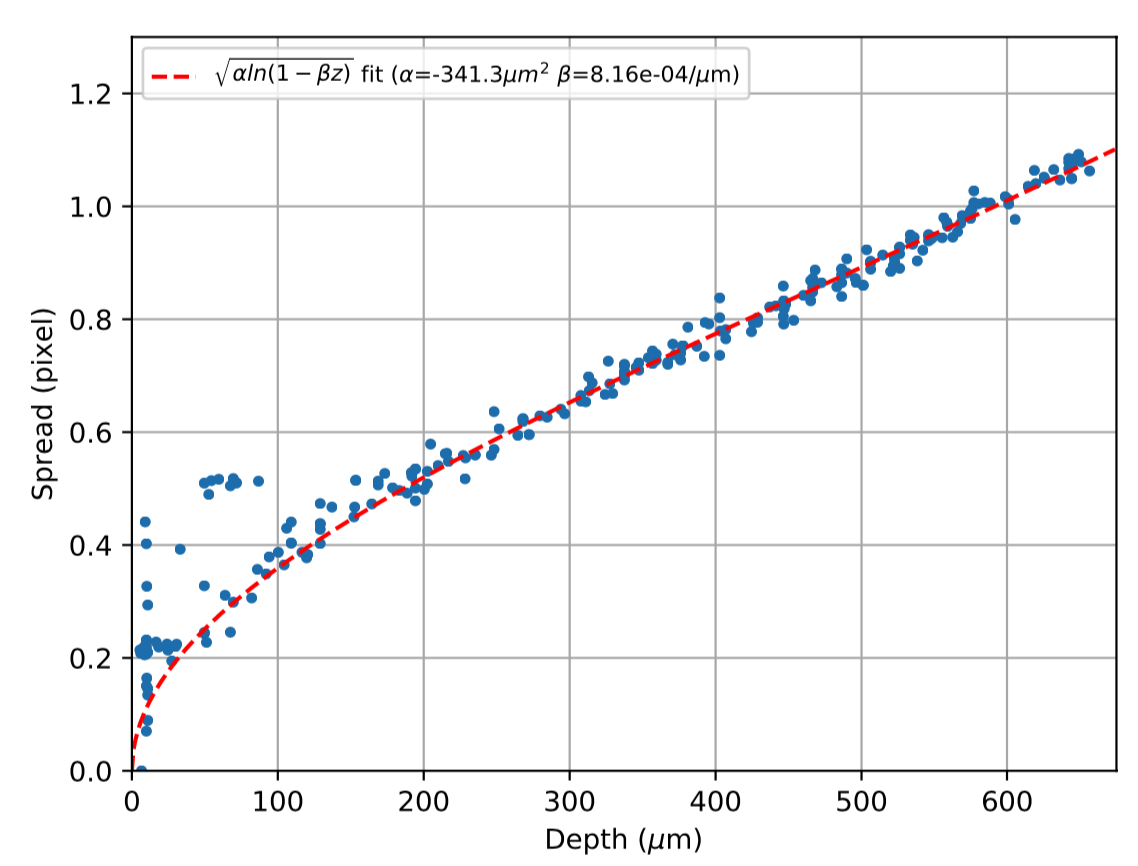}
    \caption{Data points and curve that relates the spread of the events with their interaction depth for one of the CCDs. The fitted function $\sqrt{\alpha \ln(1-\beta z)}$, where $z$ is the depth in microns, is derived from the charge transport physics in the CCD~\cite{haro2020studies}.}
    \label{fig:curve}
\end{figure}

Muon tracks are used to determine the calibration curve. 
Cosmic muons easily pass through the whole detector depth, leaving a straight track.
The electric field in the CCD volume causes the holes generated to drift perpendicularly to its front side, in the direction of the pixel collection wells. 
Therefore, the muon event observed in the image is the projection of the muon track in the plane of the CCD front-side surface. 
Figure~\ref{fig:muon1x5} shows an event of a so-called y-axis muon, whose track is perpendicular to the CCD horizontal register and is in the direction of the vertical pixel binning. 
Highlighted in the event image is a one-pixel slice, which includes one pixel in the y-direction (corresponding to five CCD pixels after the binning, or 75~$\mu$m) and all the event pixels in the x-direction. 

Figure~\ref{fig:muon1x5} also shows a plot of the charge of each muon slice and its lateral spread. The lateral spread is Gaussian with variance that depends on the time that free carriers have to diffuse laterally before being collected by the potential wells at the front of the sensor~\cite{haro2020studies}. 
This time is proportional to the depth of the ionisation location. 
The thinner side of the muon track corresponds to holes that were produced close to the CCD front side, and the thicker side, to the holes produced close to the back side. 
Due to the straight trajectory of the muon, simple trigonometry can be used to assign a depth to the lateral spread of each muon slice and compose a calibration curve. 

The spread in each muon slice was estimated by an unbiased maximum likelihood estimator described in~\cite{haro2020studies}. The resulting calibration curves, obtained separately for each CCD, show smaller spread at a given depth than in the previous CONNIE study~\cite{connie:2019}. 
Figure~\ref{fig:curve} shows the resulting curve obtained for one CCD\@. 
These size-to-depth calibration curves are used to estimate the neutrino detection efficiency for each CCD. It should be mentioned that the event size also depends on the energy deposit per pixel and that the muon curves give therefore a conservative estimate 
for low-energy deposits from neutrino interactions, as the muon energy deposits are more spread out due to the charge repulsion effect, compared with the small energy deposits by neutrinos that undergo diffusion only~\cite{haro2020studies}.

\subsection{Readout noise and single electron event stability}

\begin{figure}[tb]
\centering
\includegraphics[width=0.8\textwidth]{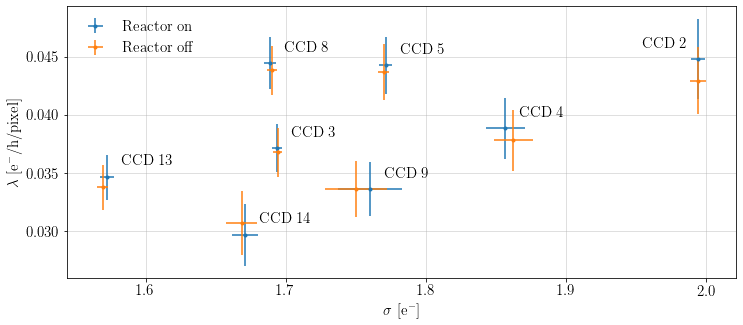}
\caption{\label{fig:noiseDC}
Readout noise (x-axis, in e$^-$) and single electron event levels (y-axis, in e$^-$/h/pix) for each CCD in reactor off and on datasets.}
\end{figure}

The main contribution to the pixel charge uncertainty is given by the readout noise (RN) added by the output amplifiers and by the spurious charge accumulated in the pixels (due to spontaneous thermal emission or dark current, Cherenkov and recombination photons~\cite{du2020sources}), the so-called single electron events (SEE)~\cite{Barak:2021crb}. 
The two effects are independent but are present in the active region of the CCD\@. 
To decouple the two, the RN is extracted directly from the overscan (unexposed) region 
by fitting a Gaussian function to the charge distribution of the pixels. 
The SEE is modeled by a Poisson distribution and is estimated using the pixels without events in the active region. 
For this, the combined probability distribution $f(E) = P(q;\lambda) * G(E; \sigma, \mu)$ is computed,
given by:
\begin{equation}
f(E) = P(q;\lambda) * G(E; \sigma, \mu)
= \frac{e^{-\lambda}}{\sqrt{2\pi}\sigma} \sum_{q=0} \frac{ \lambda^q}{q!} 
\exp\left( -\frac{(E - (\mu+g q))^2 }{2\sigma^2} \right) \,,
\label{eqn:DCeq}
\end{equation}
where $E$ is the measured pixel charge value in analog-to-digital units (ADU), $q$ runs over all possible numbers of electrons from the thermal process with a Poisson parameter $\lambda$, $g$ is the gain in ADU/e$^-$ and $\sigma$ is the RN extracted by the Gaussian fit in the overscan region. 

Both quantities are continuously monitored for all the acquired images. 
Figure~\ref{fig:noiseDC} shows the measured SEE and RN distributions for each CCD in the different reactor periods. The stability of the values between reactor on and off data is an indication that the sensors were running with similar performance in both periods.

\section{Background studies}\label{sec:backg}

\subsection{Muon and $\rm{Cu}_{\alpha}$ x-ray  rate stability}
\label{sec:bkg_stability}

Fluorescence x-rays are used to measure the stability of the natural radioactivity background rate and detector gain during reactor on and off operation.  
For this purpose we monitor the intensity of the $\rm{Cu}_{\alpha}$ peak, the most prominent background peak. 
A Gaussian function for the peak plus a constant background is fitted to the data spectra, split into groups of twenty consecutive images. 
The position of the copper peak (gain), amplitude (rate) and underlying flat background rate are checked for stability for all CCDs, and the gain and rates are compared between the reactor on and off periods. 
The chi-square per degree of freedom is computed from the sum over the CCDs of the differences between the values divided by their total uncertainty, $\chi^{2}/n=\sum (\frac{(\mu_{ON}-\mu_{OFF})}{\sigma\mu_{tot}})^{2}/n$ with $n=8$, obtaining values close to one, $\chi^{2}_{gain}/n=0.98$, $\chi^{2}_{peak-rate}/n=1.03$ and $\chi^{2}_{bkg-rate}/n=1.07$.

The muon rate is also checked for stability in the data taking period. Muons are extracted from the datasets with a Deep Convolutional Neural Network machine learning algorithm tuned for the specific hardware binned images of the 2019 data taking method. The algorithm was found to have a very high purity (above 96\%) and the resulting muon rates were found to be stationary using a Dickey-Fuller test~\cite{dickey_fuller}.

\subsection{Low-energy region studies}\label{sec:lowenergy}

\begin{figure}[tb]
    \centering
    \includegraphics[width=0.6\textwidth]{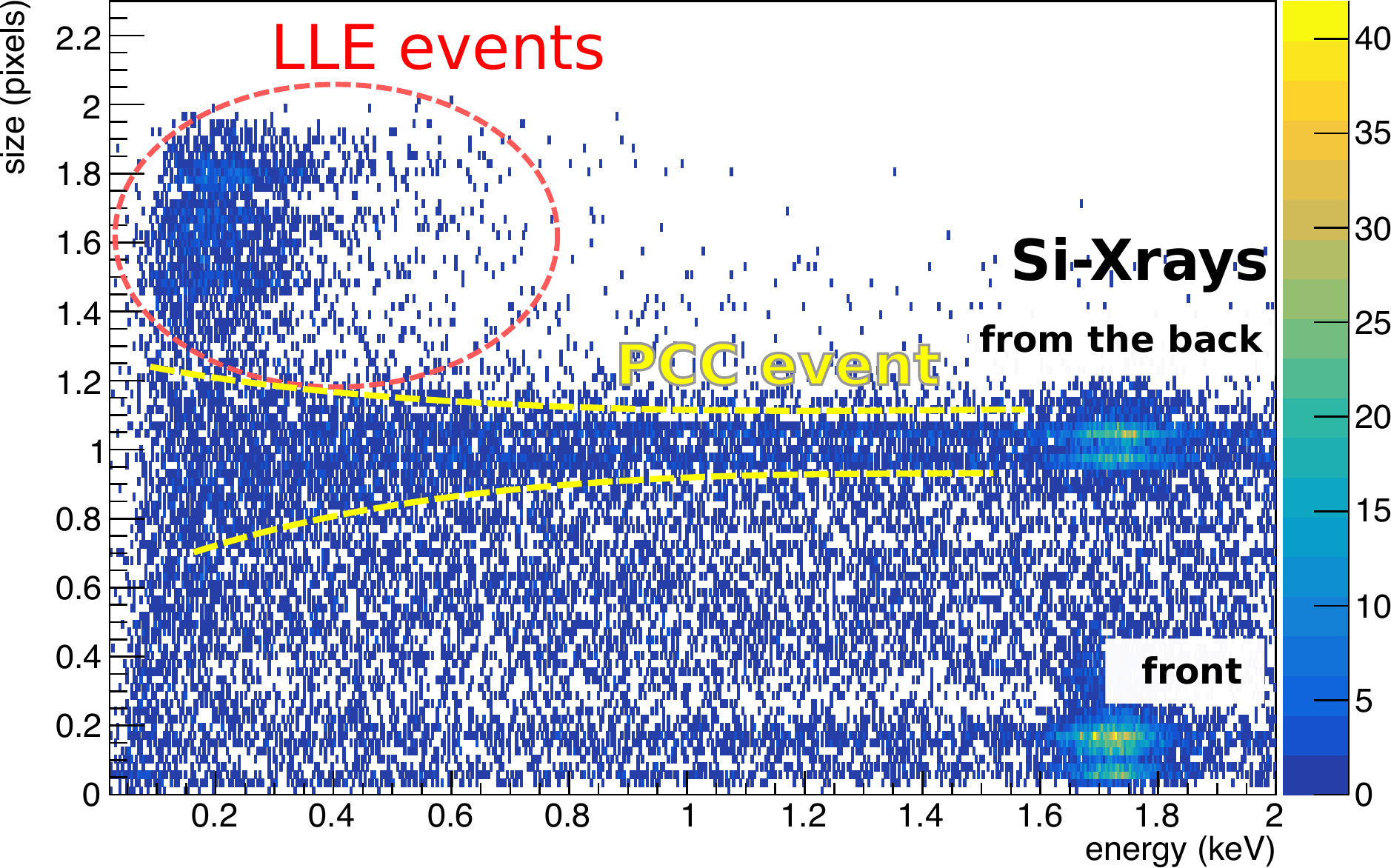}
    \caption{A two-dimensional histogram of the energy and size of low-energy events in reactor-off data. The horizontal bands are artifacts of the quantisation effect in the size fitting algorithm.}
    \label{fig:2D energy size}
\end{figure}

Two new background sources were identified in the data, that have a large impact on the event selection process in the region of interest for CE$\nu$NS interactions. 
Figure~\ref{fig:2D energy size} shows a two-dimensional histogram of the size and energy of the events after selecting an area of good pixels in the output images (excluding the edges of the sensor and any bright columns, as described in Section~\ref{sec:neutinoevents}). 
To reconstruct the size and energy of the events, similar techniques are applied as in the previous study~\cite{connie:2019}. 

Three densely populated regions are highlighted in the plot.
Around 1.7~keV there are two clusters compatible with silicon x-rays produced by fluorescence in the adjacent materials and entering by the front and back of the sensor. 
Since the attenuation depth of the photons at these energies is a few microns, most interactions occur at the very front and back of the sensor as depicted in the plot. 
There are two extra unexpected excesses of events: one called partial charge collection (PCC) events, small one-pixel events that are present in the full energy range from 0 to 2~keV; and another source called large low-energy (LLE) events, with sizes greater than 1.2 pixels and energies below 0.4~keV\@. 
Both sources have a great impact on the sensitivity to neutrino interactions at low energies. 
The following sections describe the processes that produce these events and the selection criteria to reject them.

\subsubsection{Large low-energy events}

LLE events can mimic the expected neutrino interactions, but their size is greater than that of expected events from the bulk of the sensor. 
Figure~\ref{fig:curve} shows that the maximum size for a physics induced event in the CCD is approximately 1.1 pixels. 
Two different mechanisms have been identified that may produce LLE events, based on studies of identifying their characteristic signatures in the CCD images.

The first and main type of LLE events come from the tails of very energetic events. 
Very large ionisation packets can generate tails when the charge of the pixels is transferred in the column direction. 
These tails are observed to span up to a few hundred pixels. 
This pattern does not follow the charge transfer inefficiency process observed in CCDs~\cite{janesick2001scientific}. 
At the very end of the tail, where the charge is comparable to the readout noise, the event extraction routine might identify isolated regions of low-energy pixels. 
This kind of event can be rejected by either looking at its proximity to very energetic events or by making a selection cut based on its size.

A secondary source of LLE events are charge depositions in the inactive volume of the sensor, in which some of the carriers can diffuse to the active region. 
If these depositions are produced close to the auxiliary horizontal register during the readout of the sensor, they can be erroneously seen as true depositions in the active region. 
Since these carriers are collected by a single line in the sensor, the final event shows a distinctive one-dimensional shape, generally with a broad distribution due to the diffusion in the inactive volume. 
These events can also be rejected by setting a maximum allowable size for neutrino-induced events. 

In this analysis, both types of LLE events are rejected by requiring a maximum event size, defined in Section~\ref{sec:neutinoevents}.

\subsubsection{Events from the partial charge collection layer}

\begin{figure}[tb]
    \centering
    \includegraphics[width=0.7\textwidth]{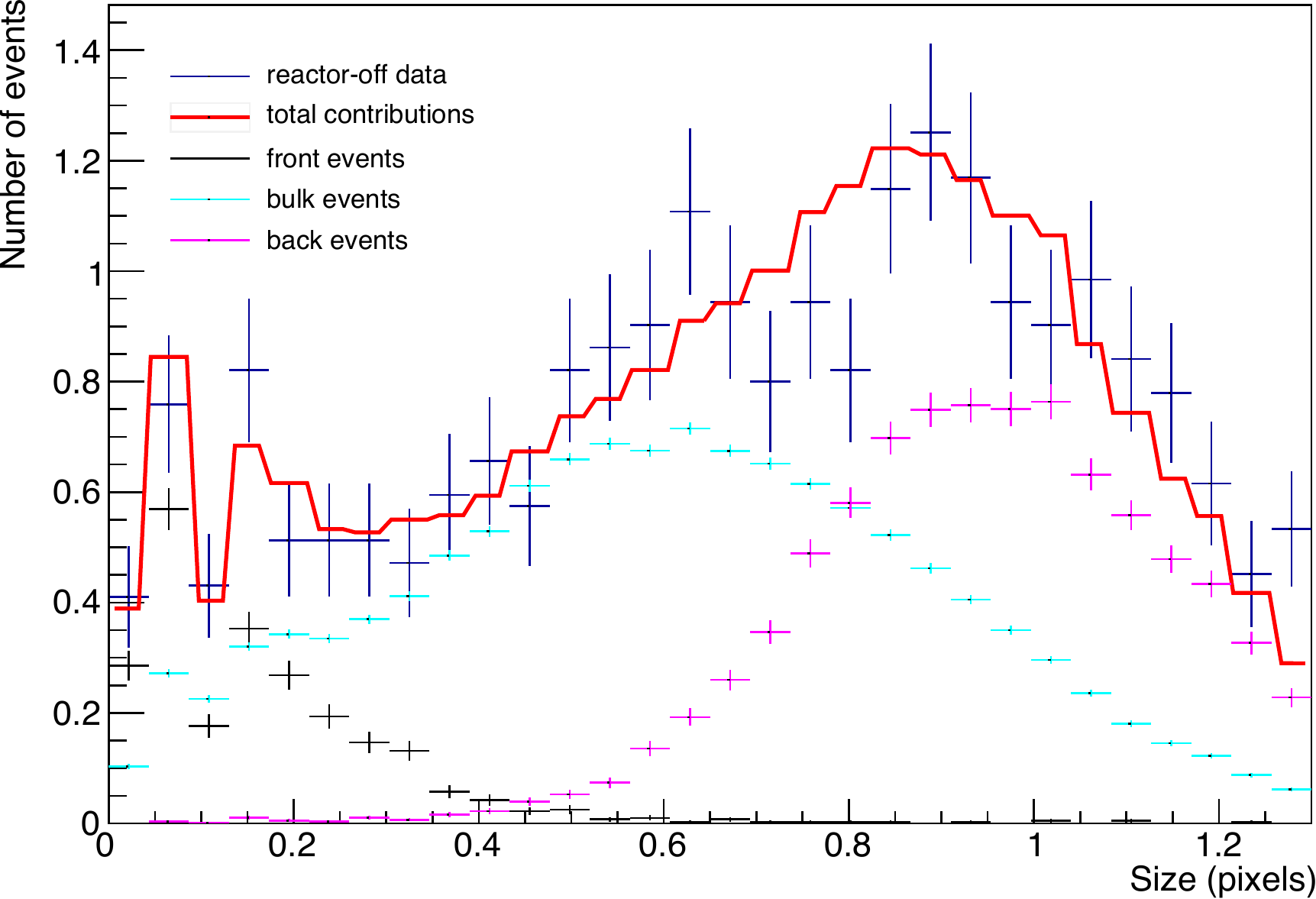}
    \caption{Size distribution of background events with energies 0.1 and 0.2\,keV from reactor-off data (blue data points), compared with simulated distributions for events from the front (black), bulk (cyan) and back (pink) regions. The red line is the sum of the simulated distributions. For events with size smaller than 0.2 pixels, most of the event charge is collected by a single pixel, so the reconstruction algorithm has little information for the size estimation and gives larger errors. This effect causes the varying structured pattern of the distributions in that region.}
    \label{fig:size_distribution}
\end{figure}

Recent studies~\cite{fernandezmoroni2020charge} have shown that extra fake events at low energy can be produced by a partial charge collection (PCC) layer in the bulk of the CCD\@.
The PCC layer is approximately 4\,$\mu$m thick at the back side of the sensor and has a high dopant concentration that prevents its depletion with the external substrate voltage. 
Free carriers are more likely to recombine in this volume before they can diffuse to the depleted silicon in the bulk of the sensor and finally be collected by the pixels. 
By this process, large charge depositions in this region by high-energy interactions can be seen as low-energy interactions if most of the free carriers recombine and only a small fraction reach the bulk of the sensor. 
Since the PCC layer is in the back of the sensor, events from it have large width, as observed in Fig.~\ref{fig:2D energy size}, which shows a clear excess of events with size close to 1 pixel. 

This effect was investigated with simulations. 
Figure~\ref{fig:size_distribution} shows the size distribution of events with energy between 0.1 and 0.2~keV collected during reactor-off operation in 2019, together with the theoretical distributions from events interacting in the very front of the sensor, uniformly distributed in the bulk, and in the back. 
These distributions are obtained by simulating low-energy interactions and transporting the free electrons until they are trapped by the pixel storage well. 
There is a good agreement between the summed contributions and the measured points. Although there is no preference in the incoming flux from the front and back at low energy, the fit reports that 10\% of events come from the front, 54\% from the bulk, and 36\% from the back, showing a clear excess from the back side. 
As in the LLE case, most of these events can be rejected in the data analysis by setting a maximum allowable size for neutrino events.

\section{Selection of neutrino candidate events}
\label{sec:neutinoevents}

The criteria applied to data to select neutrino candidates fall into three categories: temporal, geometrical and morphological selection. 
As a temporal selection, the images that show outlier values for the on-chip noise sources are removed. 
Any image with RN or SEE value 5 standard deviations above the measured mean values was excluded from further analysis. 
To be conservative, we exclude all the images obtained at the same time interval as an outlier. 
This process removes less than 0.1\% of the data sample under analysis. 

The geometrical criteria are based on the selection of good pixel regions in the sensors, and exclude all events from the edges. 
The electric field in the volume of pixels in the edge of the sensor is different from that of pixels in the center of the array due to the different border condition~\cite{janesick2001scientific}. This may change the effective volume size of those cells and therefore the charge collection efficiency and the morphology of the measured events. 
Events within 140 columns and 10 rows of the edge of the sensor were excluded from further analysis in all images. 
Moreover, CCDs can show defects in the silicon that appear as bright columns~\cite{janesick2001scientific}, in a pattern that can be different in each sensor. 
The positions of these columns are identified and no events closer than 10 pixels from them are considered.

\begin{figure}[tb]
    \centering
    \includegraphics[scale=0.78]{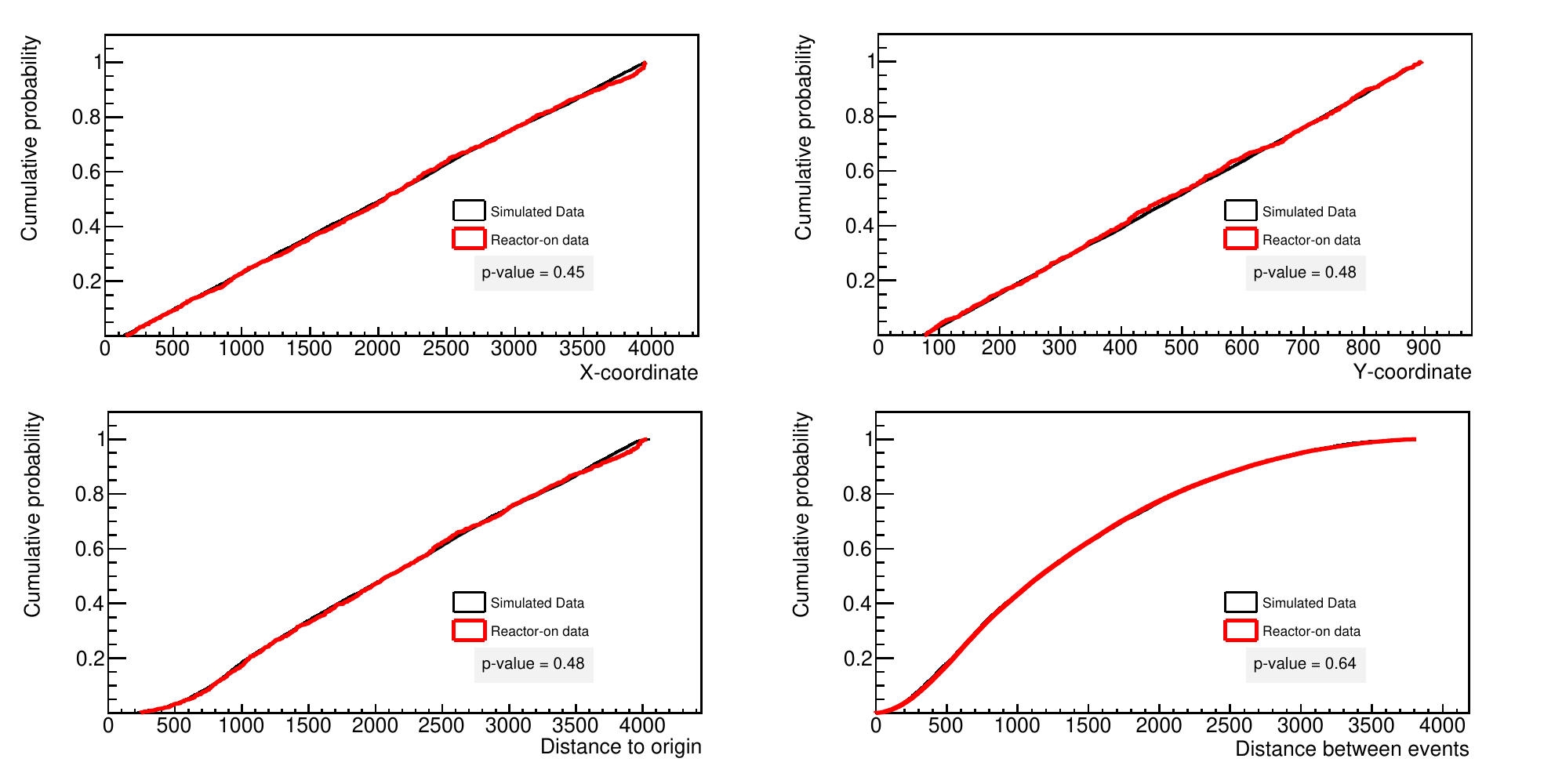}
    \caption{Cumulative distribution functions of reactor-on data and simulations for the four statistics defined in the test. The p-values are also shown.}
    \label{fig:KS1}
\end{figure}

The expected shape in the output images from a neutrino interaction is studied using the size calibration in Section~\ref{sec:size_vs_depth_calibration}. 
Since the primary ionisation volume from the neutrino scattering is expected to be much smaller than the pixel size, the final shape of the event is defined by posterior diffusion of the free carriers in the silicon. 
The calibration of this process allows to simulate events and optimise the selection criteria based on the morphology of the cluster. 
To reject the fake events produced by the on-chip noise sources, a cut on the energy of the core of the event ($E_0$) was applied, corresponding to about 4--5 times the RN, depending on the CCD\@.
Only events with $E_0>45$~eV are considered in the current analysis. 
This threshold was optimised from simulation of the on-chip noise sources to keep the rate of fake events below 10\% of the level of events measured during reactor-off operation, at low energies where the neutrino signal is expected. 
The second cut is based on the size of the event, which is required to be less than 0.95 pixels wide. 
This selection completely rejects the LLE background sources and maximises the signal-to-noise ratio for the events in the PCC layer.

\begin{figure}[tb]
    \centering
    \includegraphics[scale=0.78]{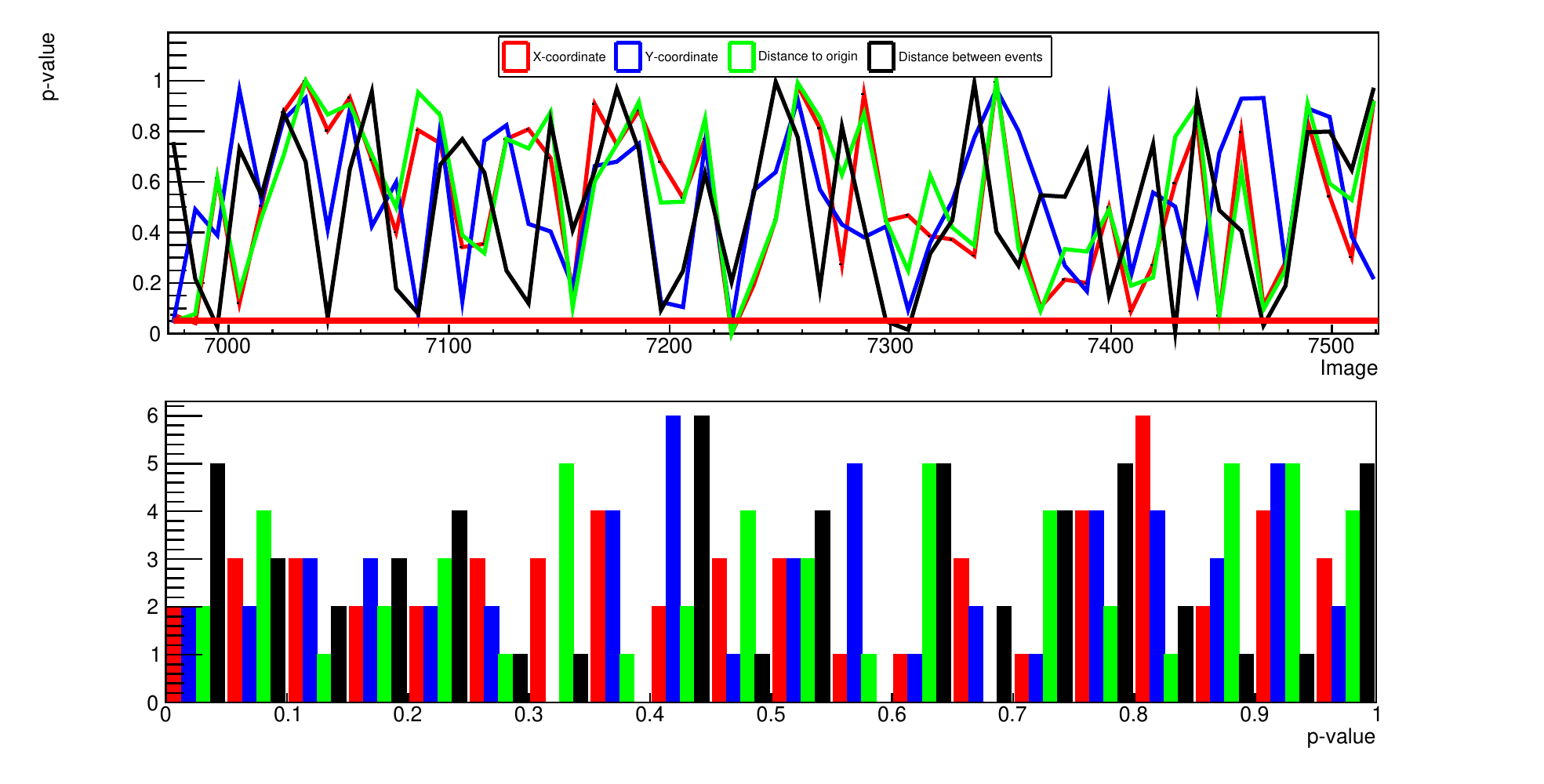}
    \caption{Time evolution of p-values for groups of ten images.}
    \label{fig:KS2}
\end{figure}

The spatial uniformity expected for neutrino events is then used to test the consistency of the data. 
Four distributions of event variables are studied: the distance to the first CCD row, the distance to the first column, the distance to the first pixel, and the distance between pairs of events.
Figure~\ref{fig:KS1} shows the distributions from data compared to the simulated samples used in the efficiency calculation, which were generated under the scenario in which the spatial distribution of neutrino events is uniform. 
A two-sample Kolmogorov-Smirnov test was applied to each pair of simulated and data distributions to determine the p-value associated with the null hypothesis that both samples are drawn from the same distribution. 
The four p-values are much larger than 0.05, the usual significance chosen to accept the null hypothesis, 
confirming the spatial distribution of data is uniform.

The same test was also applied to study the data distributions as a function of time, grouping them into sets of ten images, as shown in Fig.~\ref{fig:KS2}. 
The resulting p-values from the four considered distributions do not exhibit any trend with time. 
The fraction of the values that are smaller than 0.05 is around 5\%, corresponding to a uniformly distributed random variable.

\section{Reactor on and off spectra and CEvNS sensitivity}\label{sec:onoff}

The result of the analysis is a measurement of the neutrino interaction rate in bins of energy, which is then compared to the expected neutrino rate.  
In order to obtain the expected neutrino rates at the detector, the neutrino interaction rates predicted by the standard model are corrected for the effects of detector acceptance, selection efficiency and resolution, obtained from simulation, as well as for the quenching factor that relates the amount of ionisation measured to the nuclear recoil energy in silicon. 
The events are simulated following a uniform random distribution in the active volume, as expected from neutrino interactions and following the same procedure as in Ref.~\cite{connie:2019}.

\subsection{Detector acceptance and selection efficiency}

\begin{figure}[tb]
\centering
\includegraphics[width=.9\textwidth]{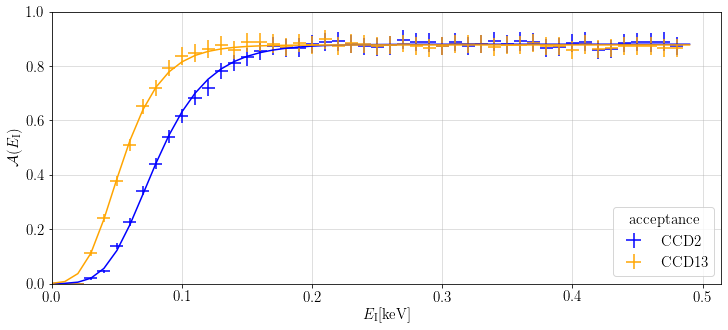}
\caption{
Extraction acceptance as a function of the ionisation energy, ${\mathcal A}(E_{\rm I})$, for the CCD with highest and lowest acceptance. The overlaid fits are performed with the function in Eq.~(\ref{eq:extrEff}).
}
\label{fig:extrAcc}
\end{figure}

The expected neutrino detection rate is determined as a function of the measured ionisation energy $E$ by applying detection effects to the neutrino rate as a function of the ionisation energy. 
It is computed by convolving this rate, corrected by the detector acceptance due to event extraction, with the Gaussian detector response and applying the efficiency of the selection cuts:
\begin{equation}
\frac{{\rm d} R }{ {\rm d} E }
= \varepsilon(E)
\int_{-\infty}^{+\infty} {\rm d}E_{\rm I}
G\left(E_{\rm I} - E - \mu(E_{\rm I}); \sigma(E_{\rm I})\right) {\mathcal A}(E_{\rm I}) \frac{{\rm d} R }{{\rm d} E_{\rm I} } \,.\label{eq:rate}
\end{equation}
The acceptance, ${\mathcal A}(E_{\rm I})$, takes into account the ionisation energies, $E_{\rm I}$, that can be reconstructed by the extraction procedure, and is computed by simulating neutrino events on top of the reactor-off images and putting them through the full processing chain.
It represents the fraction of neutrino events that can be extracted from the image for a given ionisation energy. 
The acceptance is parameterised by the function:
\begin{equation}\label{eq:extrEff}
{\mathcal A}(E_{\rm I}) = {\mathcal A}_{\rm sat} \left(
    \frac{1}{2}\tanh
        \frac{E_{\rm I} - {\mathcal A}_0}{{\mathcal A}_\sigma } + \frac{1}{2}
\right)^{10} \,,
\end{equation}
and the parameters are extracted from fits to the simulated events. 
Table~\ref{tab:fitparams} shows the parameter values obtained for each CCD\@, while the acceptance distributions for the CCDs with highest and lowest acceptances with the overlaid fits are given in Fig.~\ref{fig:extrAcc}.
The parameter ${\mathcal A}_{\rm sat}$ represents the maximum reconstruction acceptance of 87\%, which is reached at ionisation energies around 140~eV and 200~eV for the most and the least efficient CCDs, respectively.
This is a significant improvement at low energies compared to the previous analysis~\cite{connie:2019}, 
in which the maximum was reached around 500~eV. 
The improvement in reconstruction acceptance at low energies is a result of the higher signal-to-noise ratio due to the hardware binning applied at the readout stage, combined with the reduced SEE due to the shorter image exposure time of 1 hour. 

\begin{table}[tb]
\centering
\begin{tabular}{c|ccc|ccc|cc}
CCD & ${\mathcal A}_{\rm sat}$ & $\mathcal{A}_0$[eV] & $\mathcal{A}_\sigma$[eV]  & $\mu_0$[eV] & $\mu_1$[eV] & $\mu_2$[eV] & $\sigma_0$[eV] & $\sigma_1$[eV]\\
\hline
2 & 0.878 & 9.50 & 53.2 
& $-4.14$ & 3.04 & 3.83
& 33.3 & 3.31
\\
3 & 0.876 & 0.00 & 46.4 
& $-2.87$ & 2.64 & 3.07
& 28.8 & 2.32
\\
4 & 0.876 & 4.34 & 52.8 
& $-3.71$ & 0.88 & 3.73
& 30.9 & 2.80
\\
5 & 0.874 & 5.02 & 48.2 
& $-3.10$ & 2.78 & 3.37
& 30.5 & 2.74
\\
8 & 0.874 & 3.44 & 43.8 
& $-2.83$ & 3.96 & 3.44
& 29.4 & 2.51
\\
9 & 0.870 & 4.38 & 41.2 
& $-3.54$ & 2.88 & 3.79
& 29.7 & 2.47
\\
13 & 0.870 & 0.41 & 40.1 
& $-2.60$ & 1.39 & 2.94
& 28.0 & 2.31
\\
14 & 0.869 & 3.24 & 43.3 
& $-3.07$ & 0.91 & 2.95
& 29.3 & 2.74
\\
\hline
\end{tabular}
\caption{\label{tab:fitparams}
Parameter values obtained for each CCD from the fits of: the acceptance in Eq.~(\ref{eq:extrEff}) (${\mathcal A}_{max}$, $E_{0}$, $\sigma_{\mathcal A}$); the mean ionisation energy as a function of measured energy, $\mu(E_{\rm I})$ in Eq.~(\ref{eq:muFit}) ($d$, $e$, $f$, $g$); and the standard deviation of the ionisation energy as a function of the measured energy, $\sigma(E_{\rm I})$ in Eq.~(\ref{eq:sigmaFit}) ($h$, $j$, $k$).
}
\end{table}

The Gaussian convolution takes into account the shift, $\mu(E_{\rm I})$, and dispersion $\sigma(E_{\rm I})$, in the energy determination comparing the measured energy, $E$, and the ionisation energy, $E_{\rm I}$.
For this purpose, the measured energy is computed after the processing chain and compared with the simulated ionisation energy.
The mean and standard deviation for the ionisation energies are then computed as a function of the measured energy.
The convolution takes into account all ionisation energies that contribute to a given measured energy.
This step is important to properly take into account the experimental limitations in determining the ionisation energy from the measured energy.
The mean is determined from a fit to the simulated data as:
\begin{equation}\label{eq:muFit}
\mu(E_{\rm I}) = \frac{\mu_2}{E_{\rm I}^2} + \frac{\mu_1 }{E_{\rm I}} + \mu_0\,,
\end{equation}
and the standard deviation as:
\begin{equation}\label{eq:sigmaFit}
\sigma( E_{\rm I} ) = \sigma_1 \log{E_{\rm I}} + \sigma_0 \,.
\end{equation}

\begin{figure}[tb]
\centering
\includegraphics[width=.9\textwidth]{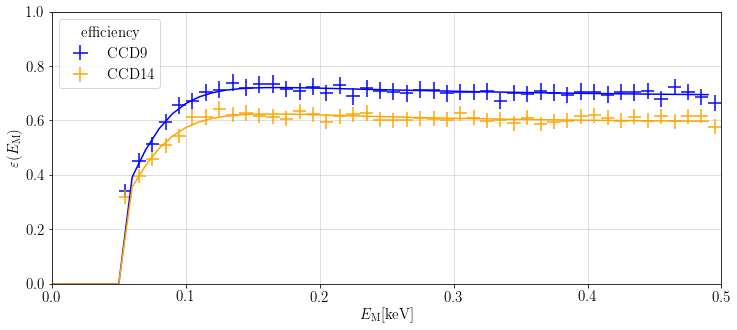}
\caption{
Selection efficiency as a function of the measured energy, $\varepsilon(E_{\rm M})$, for the CCD with highest and lowest efficiency. The overlaid fits are performed with the function in Eq.~(\ref{eq:efficiency}).
}
\label{fig:extrEff}
\end{figure}

Finally, the selection efficiency, $\varepsilon(E)$, is applied to the expected neutrino rate as a function of the measured energy to take into account the effect of the selection cuts.
The efficiency is calculated using simulations by comparing the neutrino events that pass the selection criteria to those that survive the extraction phase, and is parameterised as:
\begin{equation}\label{eq:efficiency}
    \varepsilon(E)
    =
    (\varepsilon_{\rm sat} + \varepsilon_{\delta}{\rm e}^{-E/\varepsilon_{\gamma}} ) \left(
    \frac{1}{2}\tanh
        \frac{E - \varepsilon_0}{\varepsilon_\sigma } + \frac{1}{2}
\right)^{10}\quad {\rm if}\,\, E > 50\,{\rm eV,\  else\ } \varepsilon(E)=0 
     \,,
\end{equation}
with the parameters from the fit listed in Table~\ref{tab:acceptance}. 
Figure~\ref{fig:extrEff} shows the selection efficiency as a function of energy for the most and least efficient CCDs.
The maximum efficiency of about 70\% is reached at energies of around 120~eV\@, an improvement with respect to~\cite{connie:2019}.

\begin{table}
\centering
\begin{tabular}{c|ccccc}
CCD & $\varepsilon_\sigma$[eV] & $\varepsilon_0$[eV] & $\varepsilon_{\rm sat}$ & $\varepsilon_{\gamma}$[eV] & $\varepsilon_{\delta}$\\
\hline
2 & 54.3 & $-32.9$ & 0.662 & 303 & 0.083 \\
3 & 57.1 & $-15.4$ & 0.680 & 124 & 0.225 \\
4 & 51.9 & $-20.8$ & 0.671 & 225 & 0.092 \\
5 & 59.8 & $-24.1$ & 0.682 & 185 & 0.127 \\
8 & 50.8 & $-7.82$ & 0.684 & 171 & 0.132 \\
9 & 44.1 & 0.283 & 0.685 & 236 & 0.081 \\
13 & 40.5 & 6.35 & 0.682 & 263 & 0.069 \\
14 & 47.9 & $-5.71$ & 0.593 & 158 & 0.102 \\
\hline
\end{tabular}
\caption{\label{tab:acceptance}
Parameter values for the efficiency as a function of the measured energy, $\varepsilon(E)$, obtained from a fit with Eq.~(\ref{eq:efficiency}).
}
\end{table}

\subsection{Expected neutrino rate}

\begin{table}
\centering
\begin{tabular}{ccccc}
$E$ range  & Chavarria rate & Sarkis rate & Expected 95\% CL & Observed 95\% CL \\

[eVee] & [kg$^{-1}$d$^{-1}$keV$^{-1}$] & [kg$^{-1}$d$^{-1}$keV$^{-1}$] & [kg$^{-1}$d$^{-1}$keV$^{-1}$] &  [kg$^{-1}$d$^{-1}$keV$^{-1}$] \\
\hline
50 -- 180 & 13.4 & 15.3 & 520 & 1006\\
180 -- 310 & 4.6 & 4.8 & 519 & 610\\
310 -- 440 & 1.2 & 1.3 & 504 & 422\\
440 -- 570 & 0.41 & 0.43 & 496 & 275\\
570 -- 700 & 0.15 & 0.16 & 475 & 43\\
700 -- 830 & 0.063 & 0.069 & 477 & 120\\
830 -- 960 & 0.028 & 0.031 & 485 & 719\\
\hline
\end{tabular}
\caption{
Expected neutrino rate at CONNIE in bins of measured energy in events/kg/day/keV, after applying the detector acceptance, analysis selection, and quenching factors from Chavarria~\cite{chicago_qf} and Sarkis~\cite{QF}, and expected and observed upper limits on the rates at 95\% confidence level. 
}\label{tab:expRate}
\end{table}

The reactor antineutrino flux at the detector is obtained as described in~\cite{connie:LM}, from the antineutrino spectra of fissile isotopes taken from~\cite{Vogel:1989iv} and~\cite{TEXONO:2006xds}. It is then convolved with the interaction cross-section to give the neutrino rate as a function of nuclear recoil energy, $\frac{{\rm d} R }{{\rm d} E_{\rm R} }$.
The recoil energy neutrino rate is in turn transformed into the rate as a function of ionisation energy, $\frac{{\rm d} R }{{\rm d} E_{\rm I} }$ in Eq.~\ref{eq:rate}, by applying the ionisation efficiency or quenching factor, which gives the fraction of recoil energy that causes ionisation in silicon.  
The quenching factor reflects the fact that the amount of electronic excitation produced by a recoiling ion is typically smaller than that produced by a recoiling electron of the same energy~\cite{QFSarkis2021}. 
Finally, the resulting expected neutrino rate is obtained from Eq.~\ref{eq:rate} after applying the selection efficiency, acceptance, and energy resolution. 

The expected neutrino rates as a function of measured energy are obtained separately for each CCD, and are then combined into a total rate by using the same weights as the differential (on$-$off) measured rate, given in Section~\ref{sec:onoff_rates}. 

The integrated expected neutrino rate values in bins of 130~eV are shown in Table~\ref{tab:expRate} for energies up to 1~keV\@. 
Two scenarios are considered, reflecting two different models for the quenching factor. The Chavarria quenching factor~\cite{chicago_qf}, which comes from a measurement of the ionisation efficiency
in the same type of CCD used by CONNIE and reaches energies down to 60~eVee, is included for comparison with our previous results~\cite{connie:2019}. 
The newer Sarkis model~\cite{QF,QFSarkis2021}  uses a composite solution with a model based on the extended Lindhard equation solved by a numerical method for nuclear recoil energies above 300 $\rm{eV}$, which includes the effect of the binding energy in silicon mainly relevant at low energies. 
This quenching factor shows good agreement with all of the available data for silicon and reaches energies down to 28~eVee, covering the CONNIE acceptance range at low energies. 
The Sarkis model predicts slightly higher expected detection rates, especially at the lowest energies.
The expected neutrino rates in Table~\ref{tab:expRate} have an uncertainty of around 5\% due to the reactor antineutrino flux which differs by that amount in different models~\cite{Giunti:2021kab}. This uncertainty is currently subleading to that due to the quenching factor model choice and will become relevant once the quenching factor precision is improved in future studies.

\subsection{Measured neutrino rate}\label{sec:onoff_rates}

The experimental reactor-on and off event rates in bins of energy for each CCD are obtained from the events that survive the selection during the corresponding period, in order to compare with the previously computed theoretical expected rates.
The active mass of each CCD is 5.56~g excluding the detector edges which are not considered in the analysis. The mean fiducial mass of a CCD is 4.53~g, accounting for the effect of the event size cut, which excludes interactions produced at a depth larger than about 550~$\mu$m, thus effectively reducing the active volume. 
Therefore the total experiment fiducial mass of the 8 CCDs is 36.24~g.
The total exposure time is 31.85 days with the reactor on and 28.25 days with the reactor off. 
The event rate is calculated for each CCD and period in bins of measured energy by dividing the event count by the CCD active mass, event size cut (fiducial) efficiency, exposure time and energy bin size. 
The total event rate measured by the CONNIE experiment is then obtained in bins of energy and by period, by taking the average of the 8 CCD rates, weighted by the inverse of their squared uncertainties.

\begin{figure}[tb]
\centering
\includegraphics[width=.95\textwidth]{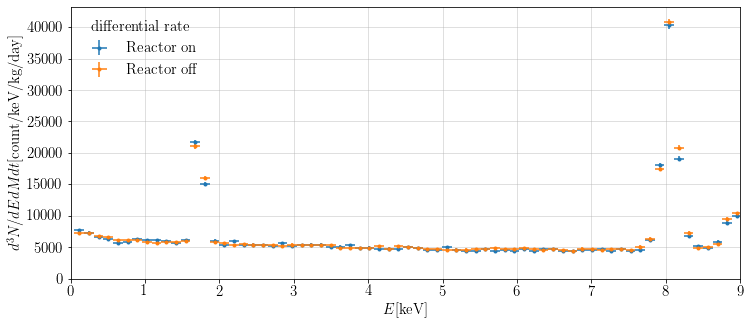}
\caption{Total event rates in bins of measured energy for reactor-on and reactor-off data. Both spectra are averaged over all CCDs.
}
\label{fig:weightedaverage}
\end{figure}

\begin{figure}[tb]
\centering
\includegraphics[width=.95\textwidth]{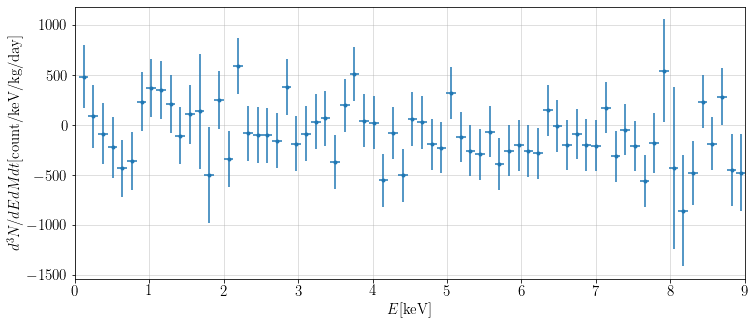}
\caption{Difference of the event rates between reactor-on and reactor-off data in bins of measured energy, averaged over CCDs.}
\label{fig:ONOFF}
\end{figure}

Figure~\ref{fig:weightedaverage} shows the total event rates for reactor-on and off periods with all the selection cuts applied. 
The two spectra show a good agreement. 
There is a substantial improvement in the background level at low energies below 1 keV, compared to the previous analysis~\cite{connie:2019}, especially in the increase of the measured spectrum towards the lowest energies. 
The small increase of the rates at very low energies that still remains can be explained by the low-energy event contribution from the PCC layers. 
Although the events are produced in the back of the sensor, the size reconstruction algorithm has worse resolution due to their low energy. 
The reconstructed sizes have a larger dispersion and some of the events pass the cut as events produced in the bulk of the sensor, as shown in Fig.~\ref{fig:2D energy size}. 
The net effect is that the selection has less rejection power at energies below 0.5 keV and some of the contribution is observed in the final spectrum. 
Further measurements in the laboratory are needed to confirm this assumption quantitatively.

\begin{figure}[tb]
\centering
\includegraphics[width=.95\textwidth]{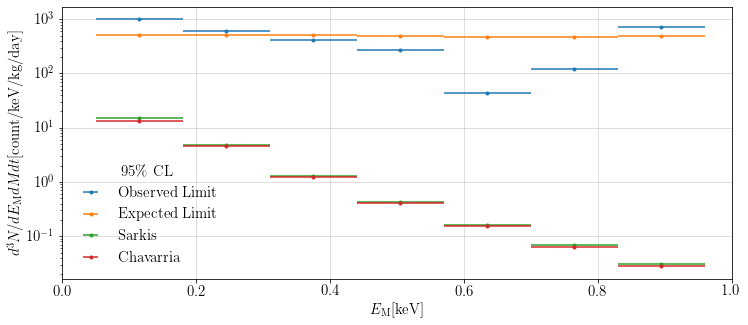}
\caption{Upper limits on the coherent neutrino interaction rate at 95\% confidence level. Observed limit (blue) and expected limit (orange) in bins of measured energy, compared with the standard model rate predictions, calculated with quenching factors from Sarkis~\cite{QF} (green) and Chavarria~\cite{chicago_qf} (red).
}
\label{fig:95CF}
\end{figure}

Figure~\ref{fig:ONOFF} shows the difference of the event rates between reactor-on and off data, resulting from the subtraction of the two spectra in Fig.~\ref{fig:weightedaverage}. 
To reduce the possible systematic effects, the subtraction is performed separately for each CCD first and then the individual results are combined optimally by the statistical uncertainty. 
The resulting difference in event rates is statistically compatible with zero.


We do not expect significant reactor-induced background, given the location of the CONNIE detector with respect to the core of the nuclear reactor. Such expectations are corroborated by a detailed study of reactor-correlated neutron backgrounds by the CONUS experiment, which is housed at 17 m from the core inside the dome of an identical nuclear reactor in Brokdorf, Germany. The performed measurements and simulations of the reactor-induced neutron fluence at the CONUS site found it to be at least an order of magnitude lower than the expected neutrino signal~\cite{Hakenmuller:2019ecb}. Considering the same reactor thermal power at the two experiments, the larger distance of CONNIE from the core and the extra material of the concrete reactor dome wall, and the comparable overall effect of the polyethylene shielding of the two experiments, any contribution from reactor-induced neutrons would be negligible compared to the expected signal and muon-induced backgrounds, and would fall below the current experimental sensitivity. The continuous measurement of the fluorescence peaks in the energy spectrum discussed in Section~\ref{sec:bkg_stability} is used currently to monitor the reactor-correlated background. Additionally, neutron detectors are being considered in the future to measure potential neutron background from the reactor.

The event rate differences between reactor-on and off data for the lowest-energy bins up to 1~keV, where the expected neutrino rates are highest, are used to set upper limits on the measured neutrino rates at CONNIE\@. 
The 95\% one-sided confidence level limit is computed from the averaged differential rate and the same averaging weights are used to combine the expected neutrino rates for each CCD\@. 
Figure~\ref{fig:95CF} shows the comparison between the expected and observed limits of the differential measured rate, and the expected rates from the standard model using the Chavarria and Sarkis quenching factors. 
The values of the limits and expected rates in each bin are given in Table~\ref{tab:expRate}. 
The observed upper limit in the lowest-energy range of $50-180$~eV is larger than the standard model rate by a factor of 66 (75) times for the Sarkis (Chavarria) quenching factor. While this result is not as strong as our previous one\,\cite{connie:2019}, it should be noted that the sensitivity of these new data is a factor of 2 better. 
The expected limit from the new data would be larger than the standard model rate by a factor of 34 (39) times for the Sarkis (Chavarria) quenching factor in the same range.
However, due to a positive fluctuation of the on-off rates the actual observed limit is worse. The opposite happened in \cite{connie:2019} where a lucky negative fluctuation of the difference between reactor-on and off spectra allowed to set a limit of about 40 times the standard model rate, while the expected limit for the previous data was around 65.

\section{Conclusion} \label{sec:conclusion}

In summary, the CONNIE experiment operated in 2019 with a readout mode of hardware binning of 5 in the vertical direction, which permits to decrease the readout noise and achieve a detection threshold energy of 50~eV\@. 
The analysis of 2.2 kg-days of 2019 data makes use of a number of new calibration tools and studies that improve the detector performance and the understanding of backgrounds. As a result, the background rate at low energies is reduced with respect to the previous analysis. 

The measured energy spectra show no excess of reactor-on data in comparison to reactor-off, yielding an upper limit at 95\% confidence level for the neutrino interaction rate in the $50-180$~eV energy range of 551~counts/keV/kg/day (expected limit) and 1055~counts/keV/kg/day (observed limit). The expected (observed) limit corresponds to 39 (75) times the standard model expectation when using the Chavarria quenching factor, and 34 (66) times the expected rate with the more recent Sarkis quenching factor. 
The possible combination of these results with the previous limit, obtained from 2016-2018 CONNIE data~\cite{connie:2019}, would require combining strongly-correlated measurements and predictions for detectors with different efficiencies, and is left for future work.

The perspective to further lower the detection threshold can be achieved by employing the recently developed skipper CCD~\cite{Tiffenberg:2017aac} sensors, in which the readout stage is modified to allow multiple non-destructive sampling of the same pixel. 
This decreases the noise down to sub-electron levels and makes them capable of counting the number of electrons in each pixel. 
The CONNIE detector was updated with two skipper-CCD sensors and their dedicated readout electronics~\cite{Cancelo:2021qmc} in mid 2021 and is currently commissioning the new setup, performing background measurements and detector characterisation since July 2021. 
This is the first step in the direction of a future skipper-CCD experiment of larger mass that will be able to achieve the increased sensitivity necessary to detect the coherent scattering of reactor antineutrinos.

\acknowledgments
We thank the Silicon Detector Facility team at Fermi National Accelerator Laboratory for being the host lab for the assembly and testing of the detectors components used in the CONNIE experiment. The CCD development work was supported in part by the Director, Office of Science, of the U.S. Department of Energy under Contract No.~DE-AC02-05CH11231.
We express gratitude to Eletrobras Eletronuclear, and especially Ilson Soares and Livia Werneck Oliveira, for access to the Angra 2 reactor site, infrastructure and the support of their personnel to the CONNIE activities. 
We thank Ronald Shellard for supporting the experiment and Marcelo Giovani for his IT support. We acknowledge the support from the  Brazilian Ministry for Science, Technology, and Innovation and the Brazilian funding agencies FAPERJ (grants E-26/110.145/2013, E-26/210.151/2016, E-26/010.002216/2019, E-26/210.079/2020), CNPq (grant 437353/2018-4), and FINEP (RENAFAE grant 01.10.0462.00); and Mexico’s CONACYT (grant No. 240666) and DGAPA-UNAM (PAPIIT grant IT100420). This work made use of the CHE cluster, managed and funded by COSMO/CBPF/MCTI, with financial support from FINEP and FAPERJ, and operating at the Javier Magnin Computing Center/CBPF.

\bibliographystyle{JHEP}

\bibliography{CONNIEbib}

\providecommand{\href}[2]{#2}\begingroup\raggedright\begin{thebibliography}{10}

\bibitem{Freedman:1973yd}
D.~Z. Freedman, {\it {Coherent Neutrino Nucleus Scattering as a Probe of the
  Weak Neutral Current}},  {\em Phys. Rev. D} {\bf 9} (1974) 1389--1392.

\bibitem{Papoulias:2019}
D.~K. Papoulias et~al., {\it {Recent probes of standard and non-standard
  neutrino physics with nuclei}},  {\em Front. in Phys.} {\bf 7} (2019) 191,
  [\href{http://arxiv.org/abs/1911.00916}{{\tt arXiv:1911.00916}}].

\bibitem{Coherent}
{\bf COHERENT} Collaboration, D.~Akimov et~al., {\it {Observation of Coherent
  Elastic Neutrino-Nucleus Scattering}},  {\em Science} {\bf 357} (2017),
  no.~6356 1123--1126, [\href{http://arxiv.org/abs/1708.01294}{{\tt
  arXiv:1708.01294}}].

\bibitem{COHERENT:2020iec}
{\bf COHERENT} Collaboration, D.~Akimov et~al., {\it {First Measurement of
  Coherent Elastic Neutrino-Nucleus Scattering on Argon}},  {\em Phys. Rev.
  Lett.} {\bf 126} (2021), no.~1 012002,
  [\href{http://arxiv.org/abs/2003.10630}{{\tt arXiv:2003.10630}}].

\bibitem{TEXONO:2005fmk}
{\bf TEXONO} Collaboration, B.~Xin et~al., {\it {Production of electron
  neutrinos at nuclear power reactors and the prospects for neutrino physics}},
   {\em Phys. Rev. D} {\bf 72} (2005) 012006,
  [\href{http://arxiv.org/abs/hep-ex/0502001}{{\tt hep-ex/0502001}}].

\bibitem{MagneticMoment}
O.~G. Miranda, D.~K. Papoulias, M.~T\'ortola, and J.~W.~F. Valle, {\it {Probing
  neutrino transition magnetic moments with coherent elastic neutrino-nucleus
  scattering}},  {\em JHEP} {\bf 07} (2019) 103,
  [\href{http://arxiv.org/abs/1905.03750}{{\tt arXiv:1905.03750}}].

\bibitem{Millicharge}
A.~Parada, {\it {Constraints on neutrino electric millicharge from experiments
  of elastic neutrino-electron interaction and future experimental proposals
  involving coherent elastic neutrino-nucleus scattering}},  {\em Adv. High
  Energy Phys.} {\bf 2020} (2020) 5908904,
  [\href{http://arxiv.org/abs/1907.04942}{{\tt arXiv:1907.04942}}].

\bibitem{Kosmas:2017zbh}
T.~S. Kosmas, D.~K. Papoulias, M.~Tortola, and J.~W.~F. Valle, {\it {Probing
  light sterile neutrino signatures at reactor and Spallation Neutron Source
  neutrino experiments}},  {\em Phys. Rev. D} {\bf 96} (2017), no.~6 063013,
  [\href{http://arxiv.org/abs/1703.00054}{{\tt arXiv:1703.00054}}].

\bibitem{WeakAngle}
B.~C. Ca\~nas, E.~A. Garc\'es, O.~G. Miranda, and A.~Parada, {\it {Future
  perspectives for a weak mixing angle measurement in coherent elastic neutrino
  nucleus scattering experiments}},  {\em Phys. Lett. B} {\bf 784} (2018)
  159--162, [\href{http://arxiv.org/abs/1806.01310}{{\tt arXiv:1806.01310}}].

\bibitem{WeakMixingAngle}
G.~Fernandez-Moroni, P.~A.~N. Machado, I.~Martinez-Soler, Y.~F. Perez-Gonzalez,
  D.~Rodrigues, and S.~Rosauro-Alcaraz, {\it {The physics potential of a
  reactor neutrino experiment with Skipper CCDs: Measuring the weak mixing
  angle}},  {\em JHEP} {\bf 03} (2021) 186,
  [\href{http://arxiv.org/abs/2009.10741}{{\tt arXiv:2009.10741}}].

\bibitem{Proceedings:2019qno}
P.~S. Bhupal~Dev et~al., {\it {Neutrino Non-Standard Interactions: A Status
  Report}},  vol.~2, p.~001, 2019.
\newblock \href{http://arxiv.org/abs/1907.00991}{{\tt arXiv:1907.00991}}.

\bibitem{Farzan:scalar}
Y.~Farzan, M.~Lindner, W.~Rodejohann, and X.-J. Xu, {\it {Probing neutrino
  coupling to a light scalar with coherent neutrino scattering}},  {\em JHEP}
  {\bf 05} (2018) 066, [\href{http://arxiv.org/abs/1802.05171}{{\tt
  arXiv:1802.05171}}].

\bibitem{CONUS:2020skt}
{\bf CONUS} Collaboration, H.~Bonet et~al., {\it {Constraints on Elastic
  Neutrino Nucleus Scattering in the Fully Coherent Regime from the CONUS
  Experiment}},  {\em Phys. Rev. Lett.} {\bf 126} (2021), no.~4 041804,
  [\href{http://arxiv.org/abs/2011.00210}{{\tt arXiv:2011.00210}}].

\bibitem{MINER}
{\bf MINER} Collaboration, G.~Agnolet et~al., {\it {Background Studies for the
  MINER Coherent Neutrino Scattering Reactor Experiment}},  {\em Nucl. Instrum.
  Meth. A} {\bf 853} (2017) 53--60,
  [\href{http://arxiv.org/abs/1609.02066}{{\tt arXiv:1609.02066}}].

\bibitem{RED:2012hpm}
{\bf RED} Collaboration, D.~Y. Akimov et~al., {\it {Prospects for observation
  of neutrino-nuclear neutral current coherent scattering with two-phase Xenon
  emission detector}},  {\em JINST} {\bf 8} (2013) P10023,
  [\href{http://arxiv.org/abs/1212.1938}{{\tt arXiv:1212.1938}}].

\bibitem{connie:2019}
{\bf CONNIE} Collaboration, A.~Aguilar-Arevalo et~al., {\it {Exploring
  low-energy neutrino physics with the Coherent Neutrino Nucleus Interaction
  Experiment}},  {\em Phys. Rev. D} {\bf 100} (2019), no.~9 092005,
  [\href{http://arxiv.org/abs/1906.02200}{{\tt arXiv:1906.02200}}].

\bibitem{connie:LM}
{\bf CONNIE} Collaboration, A.~Aguilar-Arevalo et~al., {\it {Search for light
  mediators in the low-energy data of the CONNIE reactor neutrino experiment}},
   {\em JHEP} {\bf 04} (2020) 054, [\href{http://arxiv.org/abs/1910.04951}{{\tt
  arXiv:1910.04951}}].

\bibitem{connie:2016}
{\bf CONNIE} Collaboration, A.~Aguilar-Arevalo et~al., {\it {Results of the
  Engineering Run of the Coherent Neutrino Nucleus Interaction Experiment
  (CONNIE)}},  {\em JINST} {\bf 11} (2016), no.~07 P07024,
  [\href{http://arxiv.org/abs/1604.01343}{{\tt arXiv:1604.01343}}].

\bibitem{LBNLMSL}
``{{LBNL Micro Systems Laboratory}}.''
  \url{http://engineering.lbl.gov/microsystems-laboratory/}.

\bibitem{DES:2015wtr}
{\bf DES} Collaboration, B.~Flaugher et~al., {\it {The Dark Energy Camera}},
  {\em Astron. J.} {\bf 150} (2015) 150,
  [\href{http://arxiv.org/abs/1504.02900}{{\tt arXiv:1504.02900}}].

\bibitem{DESI:2018kpn}
{\bf DESI} Collaboration, P.~Martini et~al., {\it {Overview of the Dark Energy
  Spectroscopic Instrument}},  {\em Proc. SPIE Int. Soc. Opt. Eng.} {\bf 10702}
  (2018) 107021F, [\href{http://arxiv.org/abs/1807.09287}{{\tt
  arXiv:1807.09287}}].

\bibitem{haro2020studies}
M.~S. {Haro}, G.~{Fernandez Moroni}, and J.~{Tiffenberg}, {\it {Studies on
  Small Charge Packet Transport in High-Resistivity Fully Depleted CCDs}},
  {\em IEEE Transactions on Electron Devices} {\bf 67} (2020), no.~5
  1993--2000.

\bibitem{du2020sources}
P.~Du, D.~Egana-Ugrinovic, R.~Essig, and M.~Sholapurkar, {\it Sources of
  low-energy events in low-threshold dark matter detectors},  2020.

\bibitem{Barak:2021crb}
L.~Barak et~al., {\it {SENSEI: Characterization of Single-Electron Events Using
  a Skipper-CCD}},  \href{http://arxiv.org/abs/2106.08347}{{\tt
  arXiv:2106.08347}}.

\bibitem{dickey_fuller}
D.~A. Dickey and W.~A. Fuller, {\it {Distribution of the estimators for
  autoregressive time series with a unit root}},  {\em J. American Stat.
  Assoc.} {\bf 74} (1979), no.~366 427--431.

\bibitem{janesick2001scientific}
J.~Janesick, {\em Scientific Charge-coupled Devices}.
\newblock Press Monograph Series. Society of Photo Optical, 2001.

\bibitem{fernandezmoroni2020charge}
G.~Fernandez-Moroni, K.~Andersson, A.~Botti, J.~Estrada, D.~Rodrigues, and
  J.~Tiffenberg, {\it {Charge-Collection Efficiency in Back-Illuminated
  Charge-Coupled Devices}},  {\em Phys. Rev. Applied} {\bf 15} (2021), no.~6
  064026, [\href{http://arxiv.org/abs/2007.04201}{{\tt arXiv:2007.04201}}].

\bibitem{chicago_qf}
A.~E. Chavarria et~al., {\it {Measurement of the ionization produced by sub-keV
  silicon nuclear recoils in a CCD dark matter detector}},  {\em Phys. Rev. D}
  {\bf 94} (2016), no.~8 082007, [\href{http://arxiv.org/abs/1608.00957}{{\tt
  arXiv:1608.00957}}].

\bibitem{QF}
Y.~Sarkis, A.~Aguilar-Arevalo, and J.~C. D'Olivo, {\it {Study of the ionization
  efficiency for nuclear recoils in pure crystals}},  {\em Phys. Rev. D} {\bf
  101} (2020), no.~10 102001, [\href{http://arxiv.org/abs/2001.06503}{{\tt
  arXiv:2001.06503}}].

\bibitem{Vogel:1989iv}
P.~Vogel and J.~Engel, {\it {Neutrino Electromagnetic Form-Factors}},  {\em
  Phys. Rev. D} {\bf 39} (1989) 3378.

\bibitem{TEXONO:2006xds}
{\bf TEXONO} Collaboration, H.~T. Wong et~al., {\it {A Search of Neutrino
  Magnetic Moments with a High-Purity Germanium Detector at the Kuo-Sheng
  Nuclear Power Station}},  {\em Phys. Rev. D} {\bf 75} (2007) 012001,
  [\href{http://arxiv.org/abs/hep-ex/0605006}{{\tt hep-ex/0605006}}].

\bibitem{QFSarkis2021}
Y.~Sarkis, A.~Aguilar-Arevalo, and J.~C. D\textquoteright{}Olivo, {\it {A Study
  of the Ionization Efficiency for Nuclear Recoils in Pure Crystals}},  {\em
  Phys. At. Nucl.} {\bf 84} (2021), no.~4 590--594.

\bibitem{Giunti:2021kab}
C.~Giunti, Y.~F. Li, C.~A. Ternes, and Z.~Xin, {\it {Reactor antineutrino
  anomaly in light of recent flux model refinements}},
  \href{http://arxiv.org/abs/2110.06820}{{\tt arXiv:2110.06820}}.

\bibitem{Hakenmuller:2019ecb}
J.~Hakenm\"uller et~al., {\it {Neutron-induced background in the CONUS
  experiment}},  {\em Eur. Phys. J. C} {\bf 79} (2019), no.~8 699,
  [\href{http://arxiv.org/abs/1903.09269}{{\tt arXiv:1903.09269}}].

\bibitem{Tiffenberg:2017aac}
{\bf SENSEI} Collaboration, J.~Tiffenberg, M.~Sofo-Haro, A.~Drlica-Wagner,
  R.~Essig, Y.~Guardincerri, S.~Holland, T.~Volansky, and T.-T. Yu, {\it
  {Single-electron and single-photon sensitivity with a silicon Skipper CCD}},
  {\em Phys. Rev. Lett.} {\bf 119} (2017), no.~13 131802,
  [\href{http://arxiv.org/abs/1706.00028}{{\tt arXiv:1706.00028}}].

\bibitem{Cancelo:2021qmc}
G.~I. Cancelo et~al., {\it {Low threshold acquisition controller for Skipper
  charge-coupled devices}},  {\em J. Astron. Telesc. Instrum. Syst.} {\bf 7}
  (2021), no.~1 015001.

\end{thebibliography}\endgroup

\end{document}